\documentclass[journal=jacsat,manuscript=article]{achemso}

\usepackage[version=3]{mhchem} % Formula subscripts using \ce{}
\usepackage{hyperref}
\usepackage{graphicx}
\usepackage{amsmath}
\usepackage{amssymb}
\usepackage{float} 
\usepackage{subfiles}
\author{Chetna Taneja}
\email{chetna.taneja@students.iiserpune.ac.in}
\altaffiliation{These authors contributed equally to this work}
\author{Diptabrata Paul}
\altaffiliation{These authors contributed equally to this work}
\author{G V Pavan Kumar}
\affiliation[IISER]{Department of Physics, Indian Institute of Science Education and Research (IISER), Pune 411008, India}
\alsoaffiliation{Center for Energy Science, Indian Institute of Science Education and Research (IISER), Pune 411008, India}
\title{Experimental observation of transverse spin of plasmon polaritons in a single-crystalline silver nanowire}  

\begin{document}
%\date{\today}

\begin{abstract}
We report the experimental observation of the transverse spin and associated spin-momentum locking of surface plasmon polaritons (SPPs) excited in a plasmonic single crystalline silver nanowire (AgNW). In contrast to the SPPs excited in metal films, the electromagnetic field components of the evanescent SPP mode propagating along the long axis ($x$ axis) of the NW can decay along two longitudinal planes ($x$-$y$ and $x$-$z$ planes), resulting in two orthogonal transverse spin components ($s_z$ and $s_y$). Analysis of the opposite circular polarization components of the decaying SPP mode signal in the longitudinal plane ($x$-$y$) reveals spin dependent biasing of the signal and hence the existence of transverse spin component ($s_z$). The corresponding transverse spin density ($s_3$) in the Fourier plane reveals spin-momentum locking, where the helicity of the spin is dictated by the wave-vector components of the SPP evanescent wave. Further, the results are corroborated with three-dimensional numerical calculations. The presented results showcase how a chemically prepared plasmonic AgNW can be harnessed to study optical spins in evanescent waves, and can be extrapolated to explore sub-wavelength effects including directional spin coupling and optical nano-manipulation.
\end{abstract}

\maketitle 
Angular momentum (AM) is one of the important parameters to characterize any optical electromagnetic field \cite{PhysRevA.45.8185,https://doi.org/10.1002/lpor.200810007,Bekshaev_2011}. The AM can be further decomposed into two parts, spin angular momentum (SAM) \cite{Bekshaev_2011}, related to the circular polarization of the field and orbital angular momentum (OAM), determined by the helical wave front of the field \cite{molina2007twisted,https://doi.org/10.1002/lpor.200810007,Yao:11}. In paraxial beams SAM is aligned parallel to the wave-vector \cite{Beckshaev2008} whereas for non-paraxial optical fields such as strongly focused beams or evanescent waves, SAM can have both longitudinal and transverse components with respect to the wave-vector \cite{Bliokh2014,PhysRevLett.114.063901,PhysRevX.5.011039}. The field components spinning in the longitudinal plane leads to generation of transverse SAM, whereas the longitudinal SAM stems from field components spinning in the transverse plane \cite{BLIOKH20151}. In recent years, transverse SAM has been investigated in context of spin-orbit interactions (SOIs) in strongly focused beams and in evanescent waves \cite{PhysRevLett.99.073901,Rodr,cardano2015spin,Bliokh2015,Aiello2015,Antognozzi2016,PhysRevA.103.013520}. Specifically, for evanescent waves the helicity of the SAM component gets dictated by the decay as well as propagation direction leading to ‘spin momentum locking', similar to the electronic counterpart quantum spin-Hall effect \cite{PhysRevLett.108.120403,Bliokh1448,VanMechelen:16,Triolo2017}. Different configurations have been studied in this regard such as, surface electromagnetic waves generated due to total internal reflection at an interface \cite{PhysRevA.85.061801, doi:10.1021/acsphotonics.5b00516,Qin:20}, optical fibers \cite{Petersen67}, dielectric wave guides \cite{VanMechelen:16,Abujetas2020,MAHANKALI2020124433}, and structured guided modes \cite{Shie2018816118}. The studies reveal potential utilization of SAM in various photonic applications such as spin-directional coupling \cite{Rodr_1,PhysRevLett.120.117402,doi:10.1021/acsnano.7b07379,PhysRevA.102.033518}, optical nano-probing \cite{Neugebauer2016,PhysRevLett.120.223901,Neugebauer2019}, optical tweezers \cite{doi:10.1063/5.0015991}, and longitudinal field detection \cite{C8NR01618F}.

\begin{figure}
\includegraphics{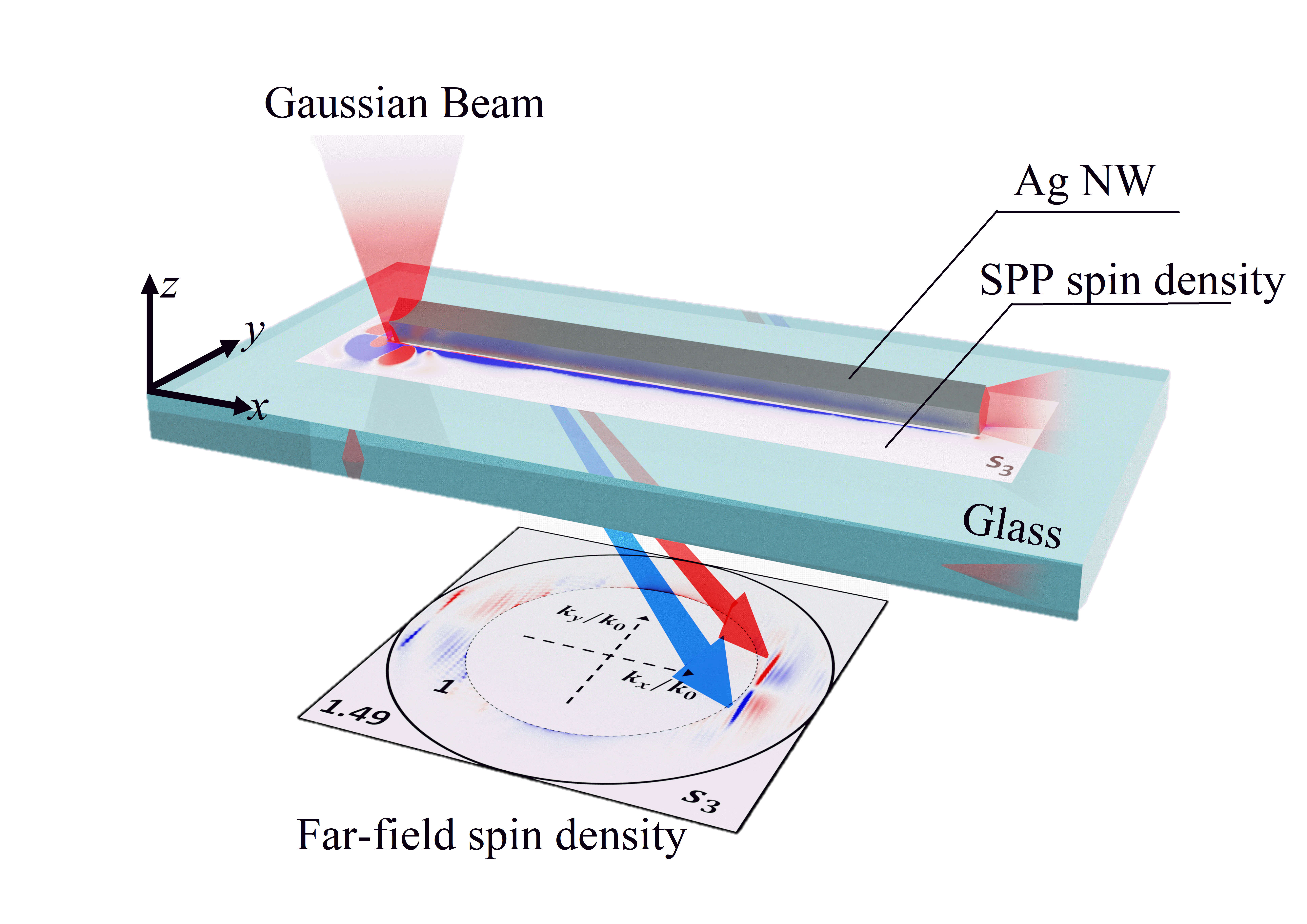}
\caption{\label{fig:1} Schematic of the measurement of the transverse spin of SPPs, excited through focused Gaussian beam in an AgNW placed on a glass substrate.}
\end{figure}

Among the evanescent surface waves such as, Dyakonov waves and Tamm waves \cite{https://doi.org/10.1002/lpor.200900050}, surface plasmon polaritons (SPPs) are of special interest due to their various emergent properties and applications in nanophotonics for broad range of wavelengths in various metallic geometries \cite{Barnes2003,ZAYATS2005131}. Of all the structures, quasi-one-dimensional single-crystalline silver nanowire (AgNW) gained particular importance because of their low loss SPP waveguiding property, leading to potential applications as an optical antenna and forms the bedrock for on-chip photonic circuitry \cite{Shegai2011,Wei2011,https://doi.org/10.1002/lpor.201200076,doi:10.1021/acsphotonics.8b01220}. In addition, uniform geometry, ease of synthesis, wide range of resonance tunability ensures ease of implementation for various nanophotonic applications on any substrate. The directional SPP propagation, unlike the case for two-dimensional metallic films or meta-surfaces, as well as electromagnetic hot-spots along the length of the AgNW during the SPP propagation have been utilized in various fields such as strong coupling \cite{Beane2018}, nonlinear optics \cite{Li2017}, remote SERS \cite{Li2020}, single photon sources \cite{Chang2007}, and cavity electrodynamics \cite{Vasista2018,https://doi.org/10.1002/adom.201900304}. In all of these cases transverse spin can play an important role in modifying the phenomena and add to a more complete description of the resultant effects such as directional coupling \cite{Gong443, PhysRevLett.123.183903}, spin-Hall effect \cite{PhysRevA.103.013520,Sharma:18}.

\begin{figure}
\includegraphics{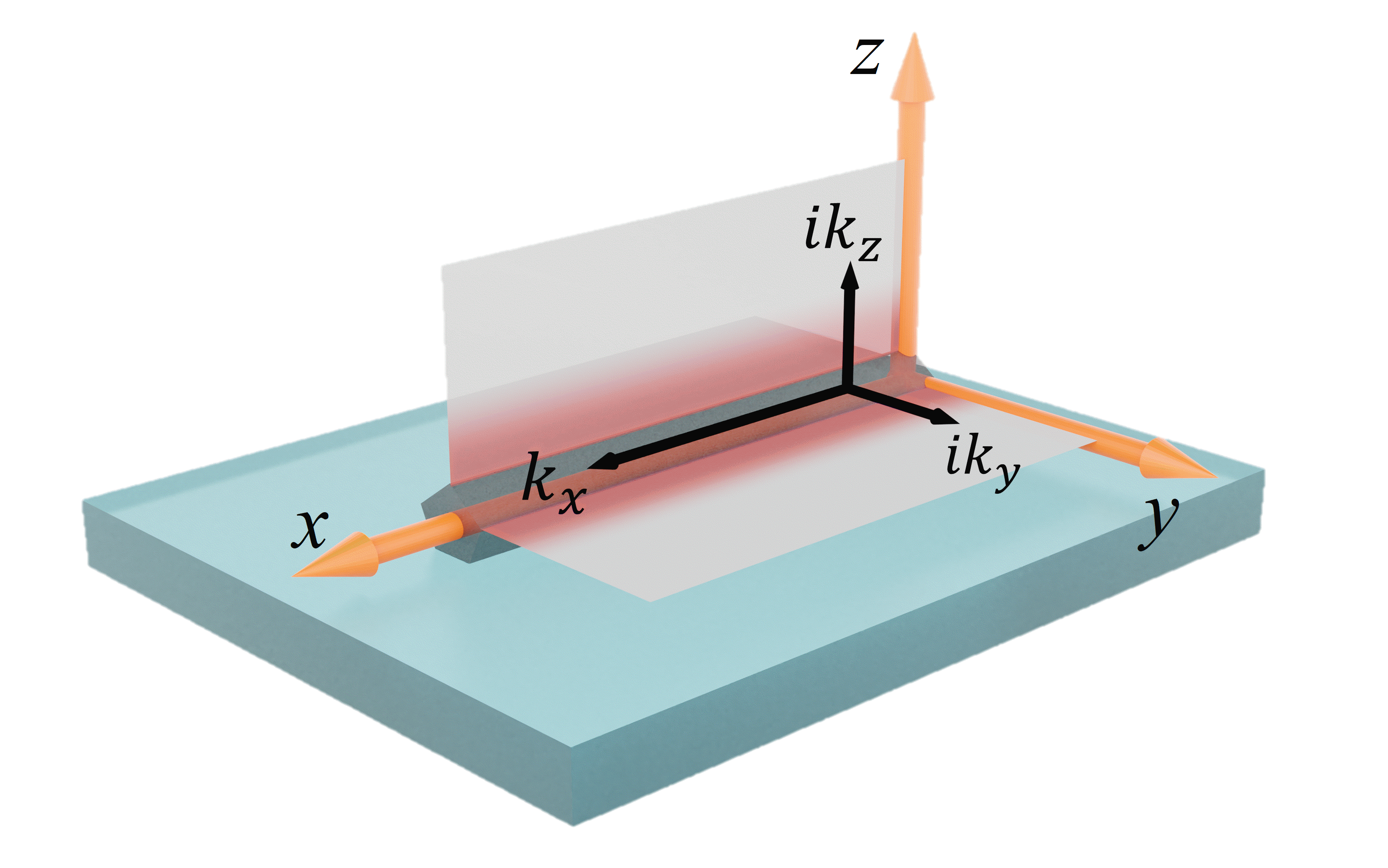}
\caption{\label{fig:2} Schematic of evanescent field of the SPPs propagating in an AgNW along $x$ axis with complex wave-vector $\mathbf{k}={k_x}\mathbf{\hat{x}}+i{k_y}\mathbf{\hat{y}}+i{k_z}\mathbf{\hat{z}}$ decaying along both $x$-$y$ ($y>0$ half space) and $x$-$z$ planes ($z>0$ half space).}
\end{figure}

Motivated by this, herein we study transverse spin of SPP field excited in a quasi-one-dimensional AgNW placed on a glass substrate as shown in Fig. \ref{fig:1}. Unlike SPPs excited on a planer metal-dielectric interface ($x$-$z$), SPPs excited in an AgNW propagating along $x$ axis can simultaneously decay along two longitudinal planes ($x$-$y$ and $x$-$z$ planes), leading to corresponding transverse spin density components ($s_z$ and $s_y$) (see Fig. \ref{fig:2}) \cite{PhysRevA.85.061801}. The experimental measurement of the transverse spin component along $z$ axis ($s_z$) relies on polarization analysis of the directly measured NW SPP mode signal excited through a tightly focused linearly polarized Gaussian beam, both in real and Fourier plane (FP). Although the directly measured SPP signal do not exhibit any biased intensity distribution with respect to the NW long axis ($x$ axis), the analyzed signal exhibit opposite bias for left circularly polarized (LCP) component with respect to the right circularly polarized (RCP) component, revealing the spin-momentum locking of SPPs. Finally, The difference between the LCP and RCP analyzed intensity distribution both in real and FP quantifies the transverse spin density distribution. The experimental observation of transverse spin of SPPs and the resultant spin-momentum locking are also corroborated with numerical calculations. 

The wave-vector corresponding to a propagating wave along any direction can only have real components, whereas for evanescent wave(s) it must also possess imaginary component(s) orthogonal to the propagation direction. For example, evanescent SPP wave propagating parallel to $x$-$y$ plane on a metal film placed on a dielectric interface can have imaginary wave-vector component in the $z>0$ direction leading to decay along $z$ axis. On the other hand, evanescent SPP waves propagating along $x$ axis on a quasi-one-dimensional metallic nano-structure can decay along both $z$ ($x$-$z$ plane) and $y$ ($x$-$y$ plane) axis. While it is well known that transverse spin density ($s_y$) exists for an evanescent wave which is decaying along $z$ axis ($x$-$z$ plane), \cite{Bliokh2014} for the current study, we investigate the properties of an evanescent wave propagating along $x$ axis and decaying along $y>0$ direction. This special case theoretically mimics the in plane ($x$-$y$ plane, $y>0$ half space) decay of SPP evanescent wave propagating along a metallic AgNW elongated along $x$ axis. 

We start with a monochromatic plane wave propagating along $x$ axis with wave-vector ($k$,0,0). The complex electric field of such a wave can be written as,
\begin{equation}
\label{eq:1}
\mathbf{E}(\mathbf{r})=E_0\frac{\mathbf{\hat{y}+\alpha\hat{z}}}{\sqrt{1+|\alpha|^2}}e^{ikx}
\end{equation}
Here, $E_0$ denotes the amplitude, $\mathbf{\hat{y}}$ and $\mathbf{\hat{z}}$ are unit vectors corresponding to cartesian axis with $k$ being the wave number. $\alpha$ determines the polarization state of the wave. We have omitted the factor $e^{i\omega t}$ for our analysis. The plane wave given by Eq. \ref{eq:1} can be transformed into an evanescent wave decaying along $y>0$ direction (imaginary wave-vector along $y$ direction) by rotational transformation of Eq. \ref{eq:1} by an angle $i\phi$ about $z$ axis, leading to the electric field expression (see supplementary information for the field calculation): 
\begin{equation}
\label{eq:2}
\mathbf{E}(\mathbf{r})=\frac{E_0}{\sqrt{1+|\alpha|^2}}({-i{k_y/k}\ \mathbf{\hat{x}}+{k_x/k}\ \mathbf{\hat{y}}+\alpha\ \mathbf{\hat{z}}})e^{i k_x x- k_y y}
\end{equation}
where, $k_x=k\cosh\phi$ and $k_y=k\sinh\phi$ are the components of the wave-vector $\mathbf{k}={k_x}\mathbf{\hat{x}}+i{k_y}\mathbf{\hat{y}}$ of the evanescent wave propagating in $x$-$y$ plane. The transverse spin density ($s_z$) calculated for such evanescent waves is thus given as:
\begin{equation}
\label{eq:3}
s_z \propto 2\frac{k_x k_y}{{|k|}^2}
\end{equation}

The expression suggests, for a given $k_x>0$ direction, the spin of the evanescent wave decaying in the direction of $k_y>0$ has opposite handedness with respect to that of $k_y<0$, resulting in spin-momentum locking of SPP evanescent fields. It must be noted that the transverse spin of the SPPs can be quantified in the far-field through measurement of the third Stokes parameter ($s_3$) since $s_3\propto s_z$.

The experimental implementation of the concept discussed above relies on SPP waveguiding property of a quasi-one-dimensional nano-structure, such as an AgNW. Chemically synthesized AgNWs \cite{Sun2003} having pentagonal cross-section with average diameter 350 nm and length $\sim30\ \mu$m were drop-casted on to a glass substrate (refractive index = 1.518). SPPs in the AgNW is generated by illuminating one end of the NW with a focused linearly polarized Gaussian beam with 100$\times$ 0.95 NA objective lens at wavelength $\lambda$ = 633 nm. High NA excitation of the NW end with polarization along the long axis ($x$ axis) of the NW ensures satisfying of the SPP momentum matching conditions and hence easier generation of the NW SPP modes. Elastic scattered light is collected using a 100$\times$ 1.49 NA objective lens in transmission configuration. The real plane and FP intensity distribution of collected light is projected on to a CCD using relay optics \cite{Hohenau:11,Kurvits:15}. The polarization is analysed by placing a combination of quarter wave-plate and an analyser in the detection path (See supplementary information, S1).

\begin{figure*}[t]
\includegraphics{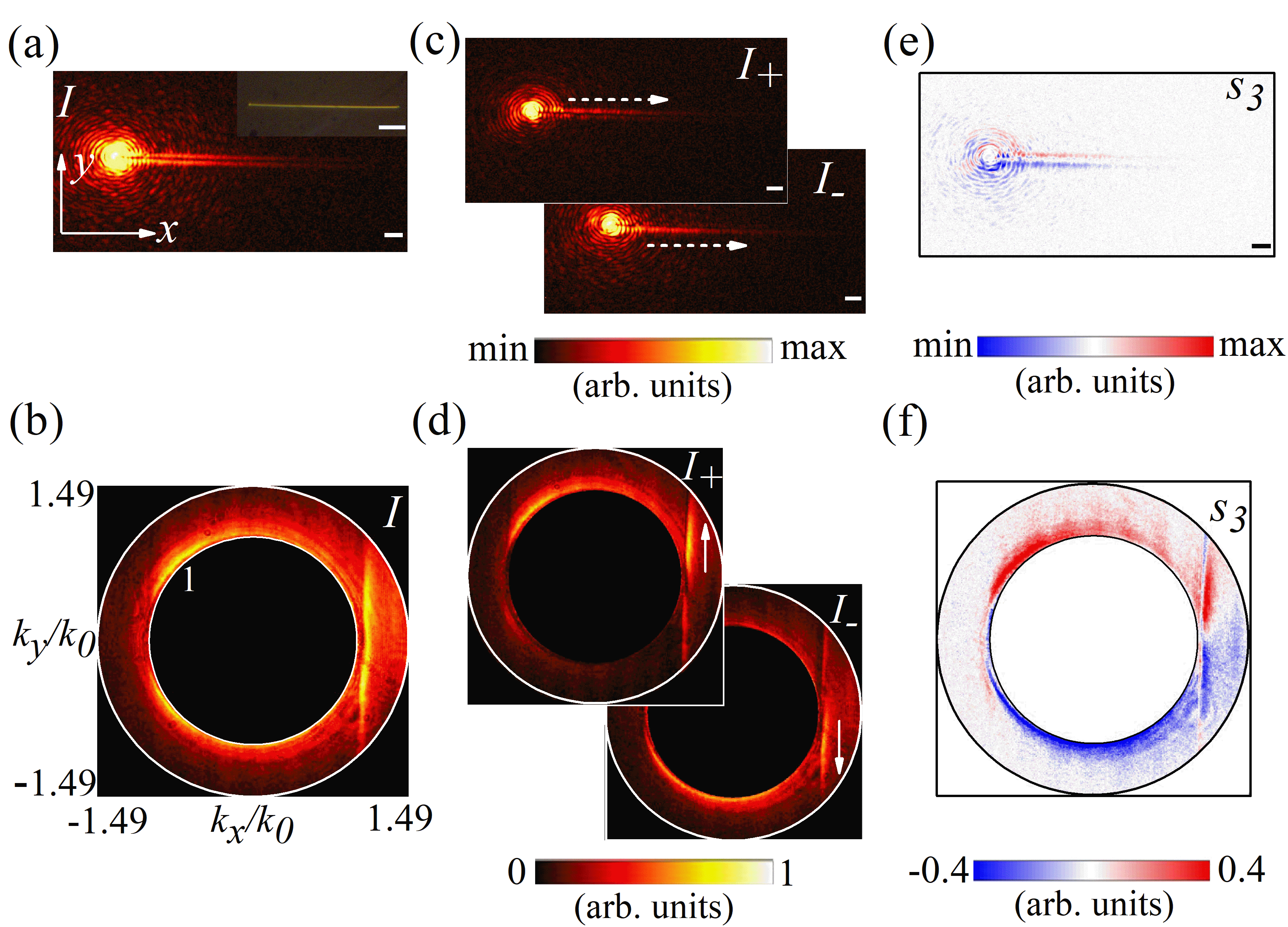}
\caption{\label{fig:3} (a) Experimentally measured real plane and the corresponding (b) Fourier plane (FP) intensity distribution of total field ($I$) of the SPP mode excited in the AgNW shown in the inset of (a). Inner and outer white circle in (b) represents collection limit of the objective lens and the region upto NA=1. A black disk at the center of (b) blocks the unscattered incident field within the inner white circle. Real plane intensity distribution corresponding to LCP ($I_+$) and RCP ($I_-$) analyzed components of the total intensity ($I$) is shown in (c). White dashed arrows indicate the bias of intensity distribution of the SPP mode. (d) The corresponding FP intensity distribution of $I_+$ and $I_-$ components, where the white arrows indicate the bias of the SPP mode line. (e) The real plane and (f) FP transverse spin density of the SPP field. The scale bar is 5 $\mu$m.}
\end{figure*}

AgNW used for our experiment can support multiple SPP modes, which can be further classified as ‘bound modes’ or ‘leaky modes’ depending on their mode refractive index (see supplementary information, S2). Evanescent near field (NF) corresponding to the bound SPP modes are majorly confined at the lower vertices of the NW and have refractive index higher than the surrounding medium and hence cannot get converted into propagating waves at the interface, leading to their bound nature. However, upon out-coupling from the distal end of the NW, they can lead to defined signatures at the far-field \cite{Song:17}.
On the other hand, the evanescent NF due of leaky SPP modes are confined around the upper vertices of the NW and have refractive index smaller than the glass substrate, allowing for their conversion into propagating waves at the air-glass interface. Thus, the far-field polarization analysis of the leaky SPP modes allows the investigation of the properties of the evanescent field \cite{https://doi.org/10.1002/lpor.201500192, Song:17}. For visualisation of the evanescent NF leaking into the substrate in far-field, we collect the scattered light through the high-refractive index substrate using oil immersion objective lens. However, imaging the scattered light using the excitation objective lens in the back-scattered geometry would not be able to capture the leaky mode NF and only the out-coupling of photons at the AgNW end or at any discontinuity along the length would be visible (See supplementary information, S3 and S4).

We start with experimental measurement of real plane intensity distribution ($I$) of a leaky SPP mode excited in an AgNW as shown in Fig. \ref{eq:3}(a) (micrograph of the NW shown in inset of \ref{fig:3}(a)). The characteristic leaky SPP mode signature can be identified as two luminous lines in $y>0$ and $y<0$ region along $x$ axis (NW long axis, excitation point being the origin).
It must be noted that the electric fields of the SPP mode also undergo exponential decay along $x$ axis due to propagation losses, dictated by the imaginary part of mode refractive index (see supplementary information, S5). Since, the length of AgNW is larger than the decay length of the SPP mode, there is no visible intensity at the distal end of the AgNW which is kept out of the imaging area.
The intensity distribution of the SPP mode exhibits symmetric distribution about $x$ axis. The corresponding FP intensity distribution is shown in Fig. \ref{fig:3}(b). The symmetric bright line along $k_y/k_0$ axis at a positive value of $k_x/k_0$ in the super-critical region ($\textrm{NA}>1$) represents the characteristic wave-vector property of the leaky SPP mode \cite{https://doi.org/10.1002/lpor.201500192}. The intensity distribution of the bright line along $k_y$/$k_0$ axis at a fixed positive value of $k_x/k_0$=1.11 (real part of refractive index of mode) indicates the confinement and decay of the mode along $y$ axis and propagation along $x$ axis in the real plane. White solid circle in FP represents the collection limit of the objective lens whereas the inner white circle indicates the region upto NA=1. The black disk at the center of FP blocks the unscattered wave-vectors lying within inner white circle. 

\begin{figure*}[t]
\includegraphics{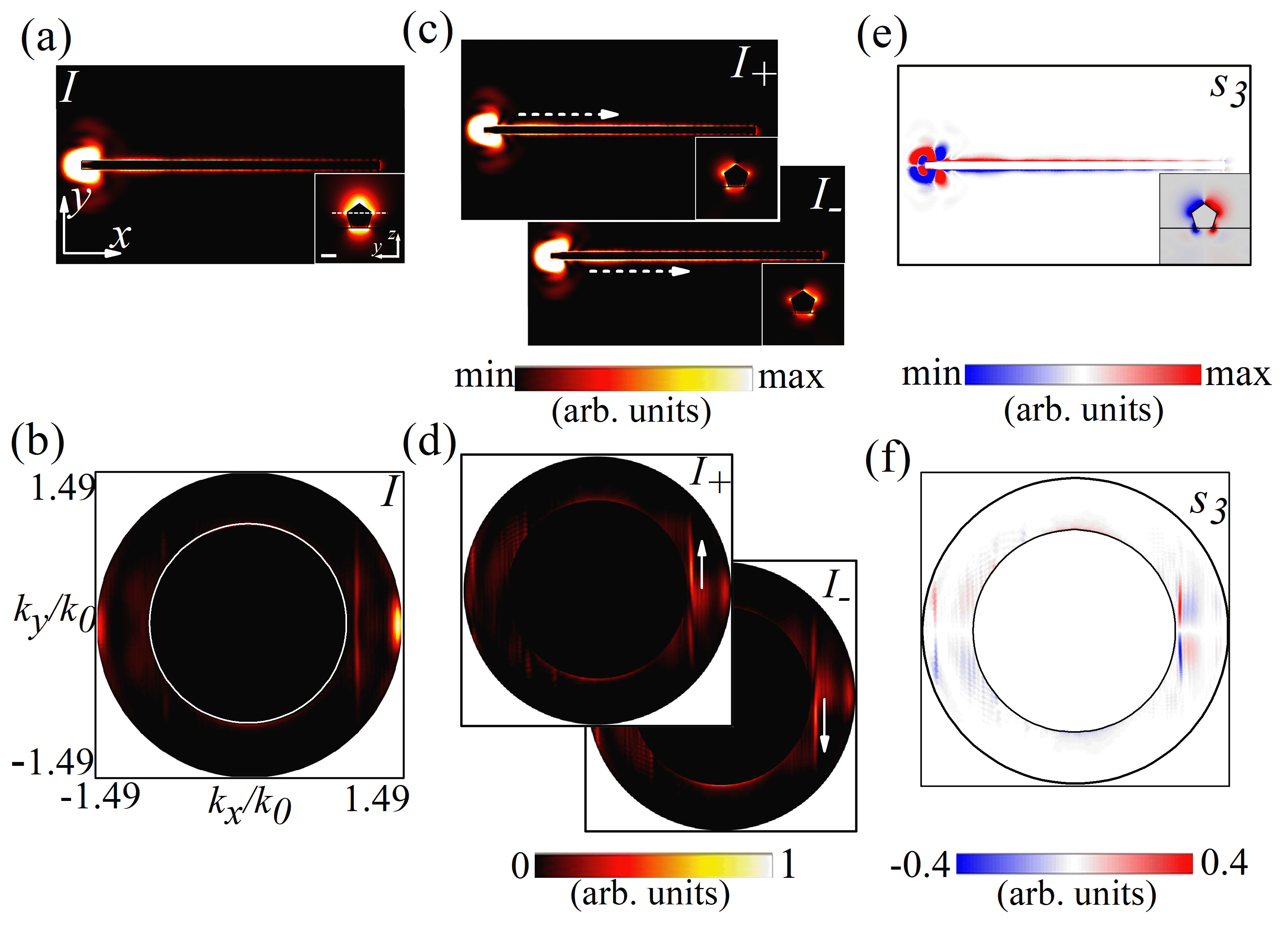}
\caption{\label{fig:4} Numerically calculated intensity distribution of the SPP mode in the real plane ($x$-$y$ plane at z = 196 nm, as indicated by white dashed line in the inset) is shown in (a) and corresponding near-field of the SPP at $y$-$z$ plane is given in the inset. (b) The corresponding FP intensity distribution. (c) Real plane intensity distribution of the SPP mode corresponding to $I_+$ and $I_-$ components and corresponding $y$-$z$ plane near-field distribution shown in the insets. White dashed arrows indicate the bias of intensity distribution of the SPP mode. (d) FP intensity distribution of $I_+$ and $I_-$ components where the white arrows indicate the bias of the SPP mode line. (e) The real plane and (f) FP transverse spin density of the SPPs field. The transverse spin distribution in the $y$-$z$ plane is shown in inset of (e).}
\end{figure*}

We now investigate the LCP ($I_+$) and RCP ($I_-$) components of the NW SPP mode and compare the corresponding real plane and FP intensity distribution with that of the total intensity ($I$). In contrast to the symmetric intensity distribution shown in Fig. \ref{fig:3}(a), the real plane $I_+$ and $I_-$ components of SPP mode as shown in Fig. \ref{fig:3}(c) exhibit bias along $y>0$ and $y<0$ luminous lines respectively, as shown by the dashed white arrows. Correspondingly, the characteristic SPP mode line in FP intensity distribution of $I_+$ and $I_-$ components shown in Fig. \ref{fig:3}(d) have bias along $k_y/k_0>0$ and $k_y/k_0<0$ respectively, shown by the solid white arrows. The differential distribution of $I_+$ and $I_-$ components both in real plane and FP indicates the existence of partial circular polarization of the SPP field. Thus, the magnitude of the transverse spin ($s_3$) in the real plane and consequently in the FP can be calculated by obtaining $s_3\propto I_+-I_-$. The real plane distribution of $s_3$ of the SPP mode obtained from the experimental data is shown in Fig. \ref{fig:3}(e). The corresponding FP distribution is given by Fig. \ref{fig:3}(f). The experimentally measured intensity values are normalized with respect to the maximum of $I$.

The existence of $s_3$ of the leaky SPP mode of the AgNW can be well understood from Eq. \ref{eq:3}. The handedness of the spin as well as its emission direction gets dictated by the propagation ($k_x/k_0$) and the decay wave-vector ($k_y/k_0$). For the SPP mode under study, the propagation direction ($k_x/k_0>0$ direction) is fixed along the long axis of the NW ($x$ axis). However, the decay wave-vector component ($k_y/k_0$) changes its sign for upper luminous line ($k_y/k_0>0$ for $y>0$) with respect to that of lower luminous line ($k_y/k_0<0$ for $y<0$). This results in $s_3>0$ for upper luminous line and $s_3<0$ for lower luminous line, i.e., opposite handedness of the transverse spin of the evanescent SPP mode decaying along opposite direction for a fixed propagation direction. The explicit dependence of the transverse spin handedness of the SPP mode excited in the AgNW on the wave-vector components indicates experimental observation of spin-momentum locking of the evanescent waves.

The experimental results have been corroborated with full wave three-dimensional finite element method (FEM) calculations. AgNW is mimicked by a geometry having pentagonal cross-section of diameter 350 nm and length 14 $\mu$m elongated along $x$ axis, with its refractive index being set to that of Ag at 633 nm wavelength. It is placed parallel to $x$-$y$ plane on a cuboid geometry mimicking a glass substrate with refractive index 1.518. Similar to the experimental configuration for excitation of SPPs, one end of the NW is illuminated with Gaussian beam at 633 nm wavelength with polarization along $x$ axis and the scattered field is calculated. The calculated NF electric field profiles were projected to the far-field by employing reciprocity arguments \cite{doi:10.1021/acsphotonics.5b00559}, since the plasmonic nano-structure interacts more strongly with the electric field of the incident electromagnetic field.

Numerically calculated intensity distribution corresponding to total field ($I$) due to leaky SPP mode in the longitudinal plane ($x$-$y$ plane at $z = 196$ nm) is shown in Fig. \ref{fig:4}(a). The plane intersects the upper vertices of the NW, as indicated by the dashed white line in the inset. The inset shows the field distribution in the transverse plane ($y$-$z$ plane). The intensity distribution indicates symmetric distribution of the generated SPP field about $x$ axis, similar to that shown in Fig. \ref{fig:3}(a). Correspondingly, the SPP mode line in the FP intensity distribution shown in Fig. \ref{fig:4}(b) exhibits symmetric intensity distribution along $k_y/k_0$ at a fixed $k_x/k_0$ value. For better comparison with the experimental results, the incident field wave-vectors upto NA=1 has been rejected in FP intensity distribution by inserting a back disk with diameter of NA=1. Along with the SPP mode line, a bright arc near the boundary of NA=1.49 and fringe pattern are also visible in the FP intensity distribution. These well studied signatures in FP distribution can be attributed to the excited bound mode of AgNW. The bright arc can be understood as the truncated Fourier transform of an excited bound mode due to the finite length of the AgNW, whereas fringe pattern represents Gibbs oscillations which originates when the length of the AgNW is comparable or shorter than the propagation length of the SPP mode \cite{Song:17,Hartmann2013}. For the current study, the main region of interest lies in the modal signature of the Leaky SPP mode.

Numerically calculated intensity distribution in the $x$-$y$ plane corresponding to $I_+$ and $I_-$ components of the total field ($I$) is shown in Fig. \ref{fig:4}(c). As in the experimentally measured case (Fig. \ref{fig:3}(c)), the SPP mode intensity distribution of $I_+$ and $I_-$ exhibit bias along $y>0$ and $y<0$ luminous lines respectively, indicated by the dashed white arrows. The intensity distribution for the SPP field at the NW edges in the $y$-$z$ plane for $I_+$ and $I_-$ components shown in the insets exhibit equivalent bias. Correspondingly, the SPP mode line in the normalised FP intensity distributions of $I_+$ and $I_-$ components exhibit bias along $k_y/k_0>0$ and $k_y/k_0<0$ respectively, as shown by the solid white arrows in Fig. \ref{fig:4}(d). The opposite bias in the intensity distribution corresponding to $I_+$ and $I_-$ components  both in $x$-$y$ plane and FP intensity distribution indicates presence of spin, which is further quantified by $s_3\propto I_+-I_-$. Fig. \ref{fig:4}(e) represents the $x$-$y$ plane $s_3$ distribution ($y$-$z$ plane shown in the inset) and the corresponding normalised FP intensity distribution is given in Fig. \ref{fig:4}(f). The simulated far-field FP distribution can be directly compared to the experimental FP intensity distribution of the plasmon mode. Although we can not directly compare the simulated scattered NF intensity distribution  with the experimental real-plane intensity distribution, the intensity biasing is clearly visible in both the cases. The numerically calculated intensity distributions corroborate our experimental observation of spin-momentum locking. The insets of the Fig. \ref{fig:4}(e) also show the existence of transverse spin in  the bound mode of the AgNW. Since, this phenomenon stems from the evanescent nature of SPP fields, both bound and leaky SPP modes supported by AgNW exhibits spin-momentum locking.

To further confirm spin-momentum locking, we numerically calculate the intensity distributions of SPP mode when the propagation direction is reversed ($k_x/k_y<0$) by changing the excitation end of the NW, shown in supplementary information S6. Since the handedness of the transverse spin of the SPP mode is dictated by the wave-vector direction, reversing the SPP propagation direction ($k_x/k_0<0$) results in the inversion of intensity distribution corresponding to $I_+$ and $I_-$ components with respect to that of $k_x/k_0>0$ case. Consequently, the handedness of the transverse spin of the SPPs decaying along $k_y/k_0>0$ (for $y>0$ luminous line) and $k_y/k_0<0$ (for $y<0$ luminous line) reverses. 

To summarize, we investigate the transverse spin ($s_z$) of SPPs propagating along $x$ axis in a quasi-one-dimensional AgNW by analyzing the circular polarization state of the collected evanescent mode signal through a high NA objective lens. Both experimentally measured and numerically calculated real plane and FP field distribution corresponding to LCP and RCP components of the total field exhibit biasing. Further, the biasing of the mode signal inverts for opposite circular polarization states indicating existence of spin-momentum locking of the evanescent wave. The existence of the transverse spin as well as the spin-momentum locking phenomena can be attributed to the decay of evanescent SPP field in the longitudinal plane ($x$-$y$ plane). Our results exhibit a direct and non-invasive measurement of the spin momentum locking as well as the corresponding transverse spin density and bypasses the complexity of fabrication based techniques \cite{Shao2018,Revah2019} or near-field nano-probing methods \cite{PhysRevLett.114.063901,Qin:20}. The results add to the emerging study of SOI due to existence of spin components in non-paraxial fields, specifically in evanescent waves. Additionally, the SOI with an AgNW will find relevance in various chip-scale nano-photonic device applications\cite{cardano2015spin}.\\

See supplementary information containing addition information on field calculation for the evanescent wave, experimental configuration, field distribution of the SPP mode, mode analysis of the experimentally excited leaky mode and numerical calculation for reversed SPP propagation.\\

\section*{AUTHORS' CONTRIBUTIONS}
CT and DP contributed equally to this work.\\

This work was partially funded by Air Force Research Laboratory grant (FA2386-18-1-4118 R\&D 18IOA118), DST Energy Science grant (SR/NM/TP-13/2016) and Swarnajayanti fellowship grant (DST/SJF/PSA-02/2017-18). CT acknowledges INSPIRE fellowship for funding. CT and DP acknowledges Rohit Chikkaraddy, Deepak K Sharma, Vandana Sharma, Sunny Tiwari, Shailendra K Chaubey for fruitful discussions.

\section*{DATA AVAILABILITY}
The data that supports the findings of this study are available within the article [and its supplementary material.

\clearpage

\begin{suppinfo}

\section{Field calculation for the evanescent wave}
Consider an electromagnetic wave propagating along $x$ axis with wave-vector ($k$,0,0). The complex electric and magnetic field components can be written as
\begin{equation}
\tag{S1}\label{eq:S1}
    \mathbf{E(r)}=\frac{{E_0}\sqrt{\mu}}{\sqrt{1
    +|\alpha|^2}} 
    \begin{pmatrix} 0 \\
    1 \\
    \alpha 
    \end{pmatrix} {e}^{ikx}
\end{equation}
\begin{equation}
\tag{S2}\label{eq:S2}
  \mathbf{H(r)}=\frac{{E_0}\sqrt{\epsilon}}{\sqrt{1
    +|\alpha|^2}} 
    \begin{pmatrix} 0 \\
    -\alpha\\
    1
    \end{pmatrix} {e}^{ikx}  
\end{equation}

Here, $E_0$ denotes the amplitude of the wave in the medium with permittivity $\epsilon$ and permeability $\mu$. $\alpha$ determines the polarization state of the wave. The polarisation state can be classified into three parameters, defined as: $\tau=\frac{1-|\alpha|^2}{1+|\alpha|^2}$, $\eta=\frac{2\mathrm{Re}\alpha}{\sqrt{1+|\alpha|^2}}$ and $\sigma=\frac{2\textrm{Im}\alpha}{1+|\alpha|^2}$. We have omitted the factor ${e}^{i\omega t}$ for our analysis.

An evanescent wave propagating along $x$ axis and decaying along $y>0$ direction (along $x$-$y$ plane) can be obtained by rotational transformation of Eq. (\ref{eq:S1}) and (\ref{eq:S2}) by an angle $i\phi$ about $z$ axis. The transformation matrix for such rotation is given as:

\begin{equation}
\tag{S3}\label{eq:S3}
    \mathbf{\hat{R}}(i\phi)= \begin{pmatrix}
\cosh{\phi} & -i\sinh{\phi} & 0\\
i\sinh{\phi} & \cosh{\phi} & 0\\
0 & 0& 1
\end{pmatrix}
\end{equation}
The transformation matrix (\ref{eq:S3}) is applied to both vector and scalar components of the fields:  $\mathbf{E(r)}\rightarrow\mathbf{\hat{R}}(i\phi)\mathbf{E}[\mathbf{\hat{R}}(-i\phi)\mathbf{r}]$ and $\mathbf{H(r)}\rightarrow\mathbf{\hat{R}}(i\phi)\mathbf{H}[\mathbf{\hat{R}}(-i\phi)\mathbf{r}]$. The application of the rotational transformation of Eq. (\ref{eq:S1}) and (\ref{eq:S2}) leads to the following expression for electric and magnetic field components: 
\begin{equation}
\tag{S4}\label{eq:S4}
\mathbf{E}(\mathbf{r})=\frac{E_0\sqrt{\mu}}{\sqrt{1+|\alpha|^2}}({-i{k_y/k}\ \mathbf{\hat{x}}+{k_x/k}\ \mathbf{\hat{y}}+\alpha\ \mathbf{\hat{z}}}){e}^{i k_x x- k_y y}
\end{equation}
\begin{equation}
\tag{S5}\label{eq:S5}
\mathbf{H}(\mathbf{r})=\frac{E_0\sqrt{\epsilon}}{\sqrt{1+|\alpha|^2}}({-i\alpha{k_y/k}\ \mathbf{\hat{x}}-\alpha{k_x/k}\ \mathbf{\hat{y}}+{\mathbf{\hat{z}}}}){e}^{i k_x x- k_y y}
\end{equation}

Here, $k_x= k\cosh{\phi}$ and $k_y= k\sinh{\phi}$ are the wave-vectors along $x$ and $y$ axis respectively. The spin density components corresponding to electric and magnetic field components given by Eq. (\ref{eq:S4}) and (\ref{eq:S5}) can be calculated  as: $\mathbf{s}=\mathbf{\Psi^\dagger}\mathbf{\hat{S}}\mathbf{\Psi}$, where $\mathbf{S}$, the Pauli spin matrices, are given by:
\begin{equation}
\tag{S6}\label{eq:S6}
\mathbf{\hat{S}_x}=(-i) \begin{pmatrix}0 & 0 & 0\\
0 & 0 & 1\\
0 & -1 & 0
\end{pmatrix}
, \mathbf{\hat{S}_y}=(-i) \begin{pmatrix}0 & 0 & -1\\
0 & 0 & 0\\
1 & 0 & 0
\end{pmatrix}
, \mathbf{\hat{S}_z}=(-i) \begin{pmatrix}0 & 1 & 0\\
-1 & 0 & 0\\
0 & 0 & 0
\end{pmatrix}
\end{equation}

Spin components associated with electric ($\mathbf{s_e}$) and magnetic ($\mathbf{s_m}$) field components for the above described evanescent wave (Eq.(\ref{eq:S4}) and (\ref{eq:S5})) are given by:
\begin{equation}
\tag{S7}\label{eq:S7}
   \mathbf{s_e}=A(\sigma\frac{k_x}{k}\mathbf{\hat{x}}-\eta\frac{k_y}{k}\mathbf{\hat{y}}+(1+\tau)\frac{k_x k_y}{k^2}\mathbf{\hat{z}})
\end{equation}
\begin{equation}
\tag{S8}\label{eq:S8}
   \mathbf{s_m}=A(\sigma\frac{k_x}{k}\mathbf{\hat{x}}+\eta\frac{k_y}{k}\mathbf{\hat{y}}+(1-\tau)\frac{k_x k_y}{k^2}\mathbf{\hat{z}})
\end{equation}

Where, A is a constant. Thus, the expression for total spin density $\mathbf{s}=\mathbf{s_e} + \mathbf{s_m}$ will be:
\begin{equation}
\tag{S9}\label{eq:S9}
   \mathbf{s}\propto2\sigma\frac{k_x}{k}\mathbf{\hat{x}}+2\frac{k_x k_y}{k^2}\mathbf{\hat{z}}
\end{equation}

The expression (Eq. (\ref{eq:S9})) suggests that, for an evanescent wave propagating along $x$ and decaying along $y>0$ direction, the spin component parallel to the propagation direction ($s_x$) is dependent on the helicity ($\sigma$), while the transverse spin along $z$ direction, $s_z$, depends on both propagation ($k_x$) and decay wave-vector ($k_y$) components of the evanescent wave. This indicates the spin-momentum locking in the evanescent wave, where the wave-vector components dictate the handedness of the transverse spin component ($s_z$). The formulation closely follows the work by K. Bliokh et al.\cite{Bliokh2014}.

\section{Experimental configuration}
\begin{figure}[h]
\includegraphics{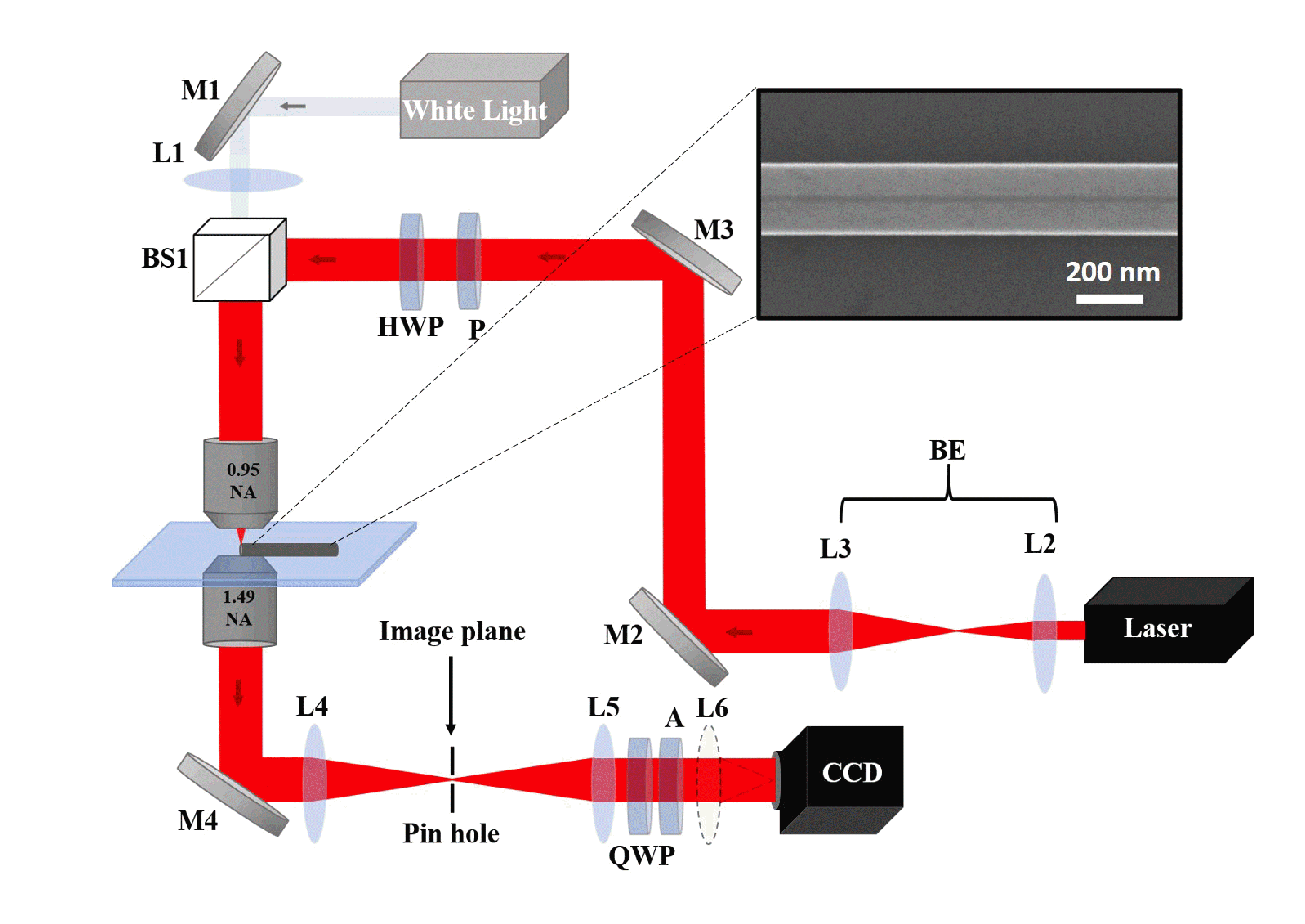}
\caption{The optical setup used for the experimental measurements. SPPs in an AgNW is excited with linearly-polarized Gaussian beam focused with $100\times$ 0.95 NA objective lens and the forward scattered light was collected using $100\times$ 1.49 NA objective lens. A polarizer (P) and half wave-plate (HWP) is used to engineer the input linear polarization state and the collected light is analyzed using a combination of quarter wave-plate (QWP) and an analyzer(A). The collected light is projected onto the CCD using relay optics for both real and Fourier plane (FP) imaging. BE is beam expander; L1-L5 are lenses; M1-M4 are mirrors and BS1 is a beam splitter. A pin hole is used to spatially filter out the region of interest in the real plane. Inset shows a scanning electron micrograph image of an AgNW section.}
\end{figure}

\section{Field distribution of the SPP modes}
\begin{figure}[H]
\includegraphics{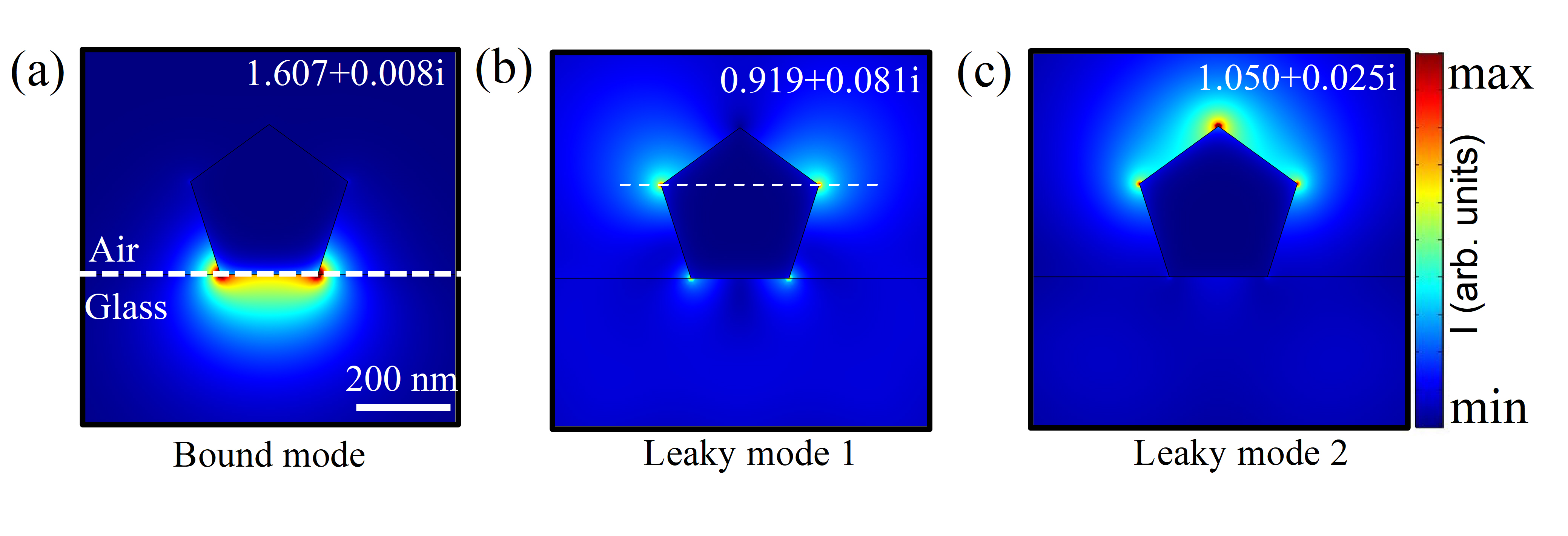}
\caption{Numerically calculated near-field of the (a) bound, (b) and (c) leaky SPP modes excited in an AgNW of 350 nm diameter placed on a glass substrate (refractive index 1.518). A bound mode having its  mode refractive index greater than the surrounding medium has its field components localized at the interface [shown in (a)], whereas for the leaky modes with refractive index lower than the substrate, the field accumulation is more around the upper vertices (white dotted line in (b)) of the AgNW \cite{https://doi.org/10.1002/lpor.201500192,Song:17}.}
\end{figure}

\section{Far-field measurement of the Plasmon modes}

\begin{figure}[H]
\centering
\includegraphics[width = 300 pt]{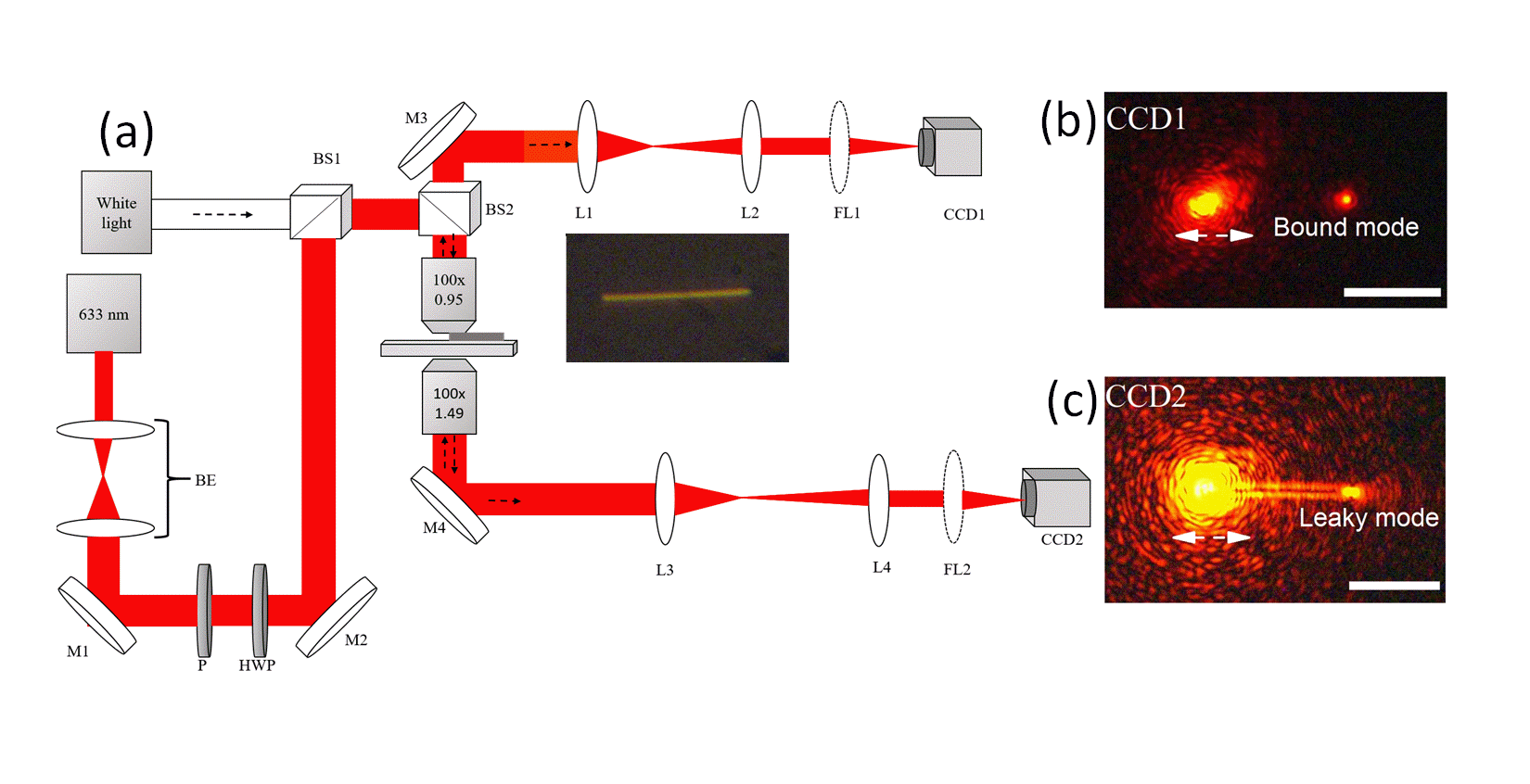}
\caption{(a) The optical setup used for the simultaneous measurement of Leaky and Bound plasmon mode.  A polarizer (P) and half wave-plate (HWP) is used to engineer the input linear polarization state. The collected light is projected onto the CCDs using relay optics for both real and Fourier plane (FP) imaging. BE is beam expander; L1-L4 are lenses; M1-M4 are mirrors and BS1-BS2 are beam splitter. Inset shows an optical image of AgNW used for these measurements.(b) and (c) Real-plane intensity distribution visualised in CCD1 and CCD2.}
\end{figure}

\begin{figure}[H]
\centering
\includegraphics[width = 300 pt]{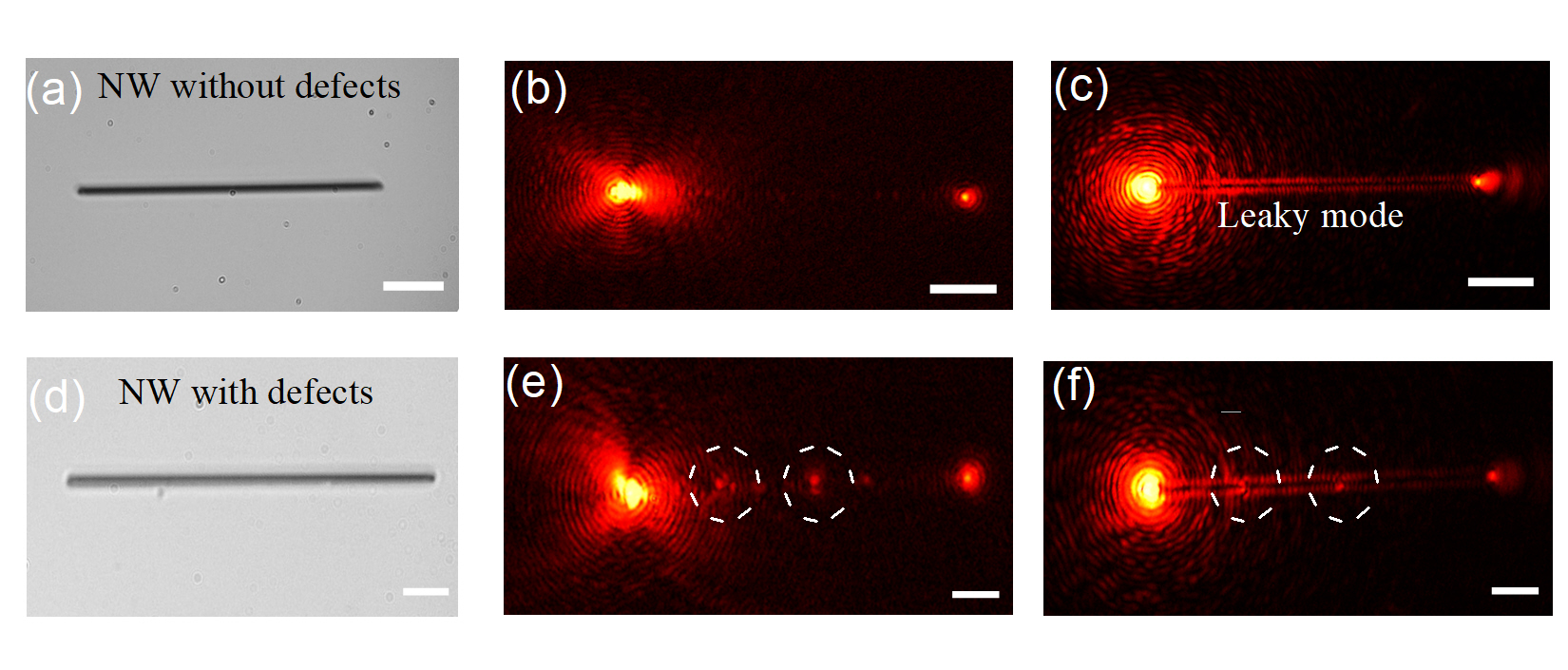}
\caption{(a) Optical bright field image of an AgNW without any defects on the surface. (b) and (c) Real-plane intensity distribution of excited bound plason mode (BPM) and leaky plasmon mode (LPM) of the AgNW by illuminating the one end of the NW with laser wavelength 633 nm polarised along the length of AgNW. (d) Optical bright field image of the AgNW with sub-wavelength defects that are not visible in the image. (e) and (f) Real-plane intensity distribution of the BPM and LPM respectively. Dashed white circles in (e) and (f) show the out-coupling of photons at the defect locations for both BPM and LPM. Plasmons were launched by exciting the end of the AgNW with 100$\times$ 0.95 NA objective lens and the scattered light from the sample plane is collected using the the oil-immersion 100$\times$ 1.49 NA objective lens through glass.}
\end{figure}

\section{Mode analysis of the experimentally excited leaky mode}
\begin{figure}[h]
\begin{center}
\includegraphics{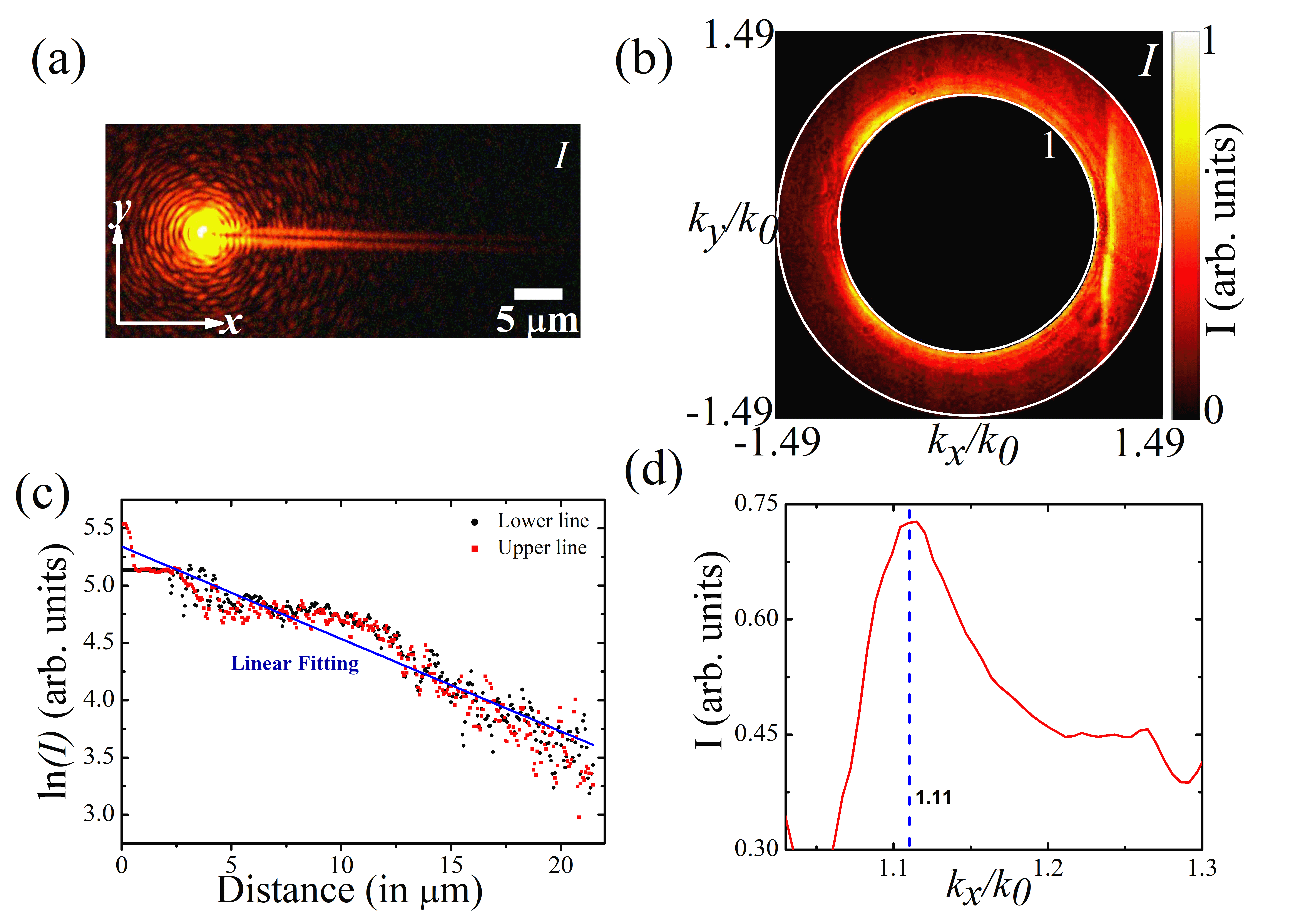}
\caption{\label{Fig:S1}(a) Real plane intensity distribution of leaky SPP mode excited in an AgNW at $\lambda$ = 633 nm. Corresponding FP intensity distribution of the mode is shown in (b). (c) Variation of natural logarithm of the Intensity distribution with the distance along the length of NW for upper (red points) and lower (blue points) luminous lines of the mode shown in (a). Solid purple line indicate the linear fit for the intensity distribution. (d) Intensity profile at $k_y/k_0=0$ and along $k_x/k_0$ axis of the Fourier plane image shown in (b).}
\end{center}
\end{figure}

The excited leaky SPP mode property ($k$ = $k_r$+$i k_i$) can be quantified by analyzing the Fig.(a) real plane  and Fig.(b) FP intensity distribution \cite{https://doi.org/10.1002/lpor.201500192}. $k_i$ can be calculated by finding out the propagation length ($L = \frac{\lambda}{4 \pi k_i}$) of the SPP mode, obtained by investigating the intensity profile ($I$) of the leaky SPP mode in real plane  along $x$ axis. Fig.(c) shows the corresponding plot of $\ln(I)$ as a function of distance from the excitation point along $x$ axis for the two luminous lines in Fig.(a). The intensity distribution is fitted with a linear equation (solid purple line) $\ln{(I)=(-\frac{1}{L})x+2\ln(A)}$, where $L$ represents the propagation length and $A$ is a fitting parameter. The value of the propagation length ($L$) from the fitting equation is found to be $L \approx 12\mu$m. $k_r$ can be obtained by investigating the FP intensity distribution in Fig.(b) along $k_x/k_0$ axis at $k_y/k_0=0$, shown in Fig.(d). The peak at the value of $k_x$/$k_0$ = $k_r$= 1.11 corresponds to the leaky SPP mode of the NW.

\section{Numerical calculation for reversed SPP propagation}

Numerically calculated intensity distribution of the SPP propagation in $x$-$y$ plane (at $z=196$ nm) in an AgNW along $k_x/k_0<0$ direction is shown in Fig.(a). While the SPP propagation direction is inverted, the decay wave-vector remains unchanged: $k_y/k_0>0$ for upper luminous line in $y>0$ region and $k_y/k_0<0$ for lower luminous line in $y<0$ region (excitation point is considered as the origin). Consequently, the circular polarization analyzed intensity distributions shown in Fig.(c) exhibit biasing (shown by dashed white arrow) opposite to that for $k_x/k_0>0$ direction SPP propagation case. The corresponding FP intensity distribution shown in Fig.(d) exhibit equivalent biasing, shown by the solid white arrows. The transverse spin density distribution ($s_3$) in $x$-$y$ plane and in FP is shown in Figs.(e) and (f) respectively.

\begin{figure}
\begin{center}
\includegraphics{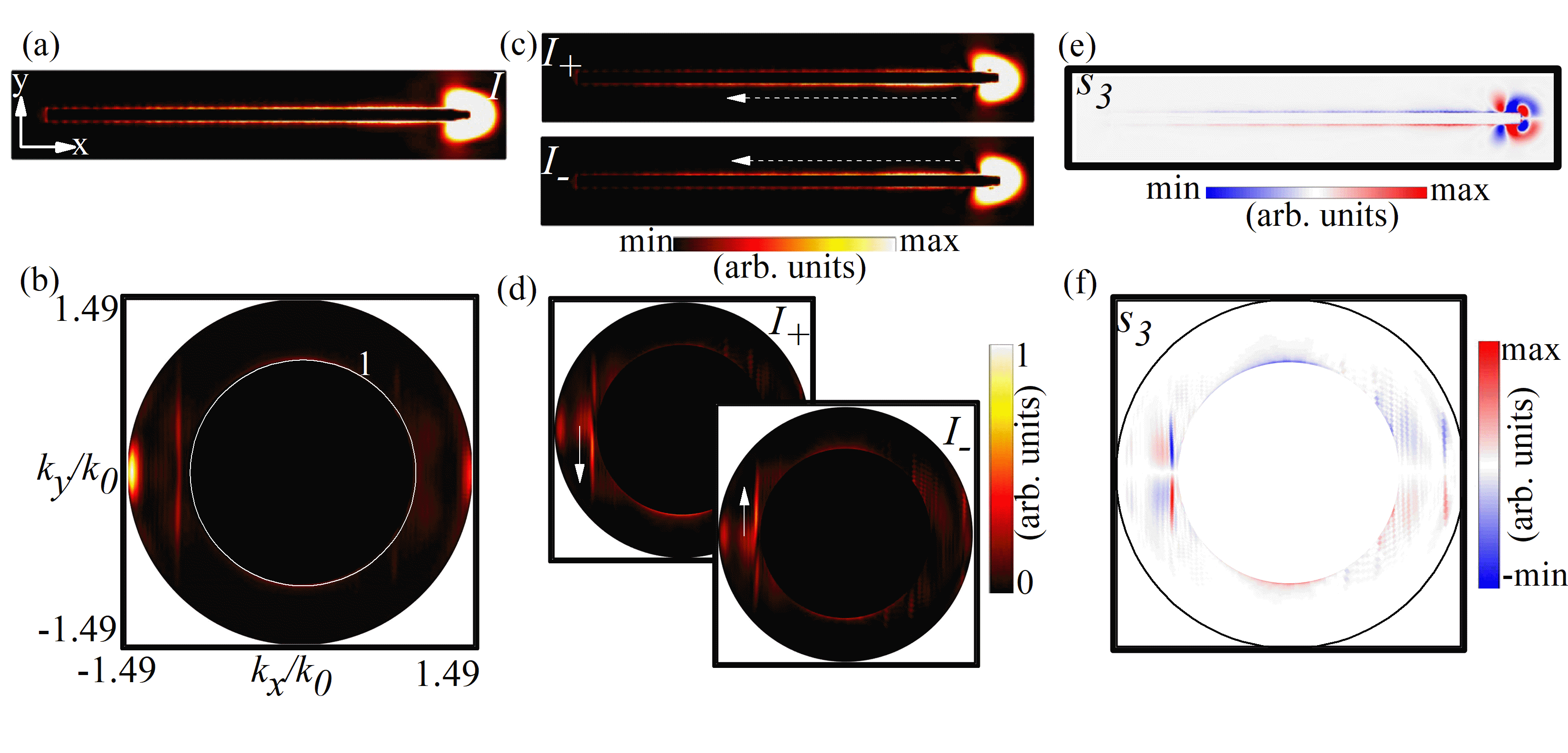}
\caption{\label{fig:4} Numerically calculated intensity distribution of the SPP mode in the real plane ($x$-$y$ plane at z = 196 nm) is shown in (a). (b) The corresponding FP intensity distribution. (c) Real plane intensity distribution of the SPP mode corresponding to $I_+$ and $I_-$ components. Dashed white arrows indicate the bias of intensity distribution of the SPP mode. (d) FP intensity distribution of $I_+$ and $I_-$ components where the solid white arrows indicate the bias of the SPP mode line. (e) The real plane and (f) FP transverse spin density of the SPPs field. }
\end{center}
\end{figure}
%\bibliography{additional}
\end{suppinfo}

\bibliography{tspin.bib}

\end{document}